\documentclass{XrU2005}
\usepackage{epsfig}

\def\ga{\mathrel{\mathchoice {\vcenter{\offinterlineskip\halign{\hfil
$\displaystyle##$\hfil\cr>\cr\sim\cr}}}
{\vcenter{\offinterlineskip\halign{\hfil$\textstyle##$\hfil\cr
>\cr\sim\cr}}}
{\vcenter{\offinterlineskip\halign{\hfil$\scriptstyle##$\hfil\cr
>\cr\sim\cr}}}
{\vcenter{\offinterlineskip\halign{\hfil$\scriptscriptstyle##$\hfil\cr
>\cr\sim\cr}}}}}

\title{The X-ray Telescope on board {\it Swift}: status and main results}
\author[1]{G. Tagliaferri}
\author[1]{S. Campana}
\author[1,2]{G. Chincarini}
\author[3]{P. Giommi}
\author[4]{G. Cusumano}
\author[5]{D.N. Burrows}
\author[5,6]{J.E. Hill}
\author[5]{J.A. Kennea}
\author[5]{J.A. Nousek}
\author[7]{J.P. Osborne}
\author[7]{P.T. O'Brien}
\author[7]{A. Wells}
\author[6]{L. Angelini}
\author[]{on behalf of the XRT team}
\affil[1]{INAF-Osservatorio Astronomico di Brera, Via Bianchi 46, 23807 Merate, Italy}
\affil[2]{Universit\`a degli Studi di Milano-Bicocca, P.za delle Scienze 3, 20126 Milano, Italy}
\affil[3]{ASI Science Data Center, Via G. Galilei, 00044 Frascati, Italy}
\affil[4]{INAF-IASF Palermo, Via U. La Malfa 153, 90146 Palermo, Italy}
\affil[5]{Pennsylvania State University, 525 Davey Lab, University Park, PA 16802, USA}
\affil[6]{NASA Goddard Space Flight Center, MD 20771, USA}
\affil[7]{University of Leicester, Department of Physics and Astronomy, Leicester, LE 17 RH, UK}
\begin{document}


\maketitle

\begin{abstract}
The X-ray Telescope (XRT), on board the {\it Swift} satellite, provides:
automated source detection and position with few arcsecond accuracy within few seconds
from target acquisition; CCD spectroscopy and imaging capability (0.2-10 keV), with
the capability of detecting a milliCrab source in about 10 seconds;
automatic adjusting of the CCD readout mode to optimize the science return as the source fades.
{\it Swift} main scientific goal is the study of gamma-ray burst (GRBs).
XRT can observe GRB afterglows over several orders of magnitude in flux.
The first results obtained during the first ten months of operation 
confirm that XRT is fully compliant with the requirements and is
providing excellent results. In particular it is detecting a very steep decay in the early
X-ray light curve of many afterglows. Often there are also strong flares superimposed 
to the X-ray light curve, probably related to the continued internal engine activity.
XRT is also localising for the first time the X-ray counterparts to short bursts.
\end{abstract}

\section{Introduction}
The first detection by the {\it Beppo}SAX satellite of a X-ray afterglow associated
with GRB\,970228 (Costa et al. 1997) revolutionised the study of the Gamma Ray Bursts
(GRBs). It was finally possible to study the counterparts of these 
elusive sources. Optical and radio afterglows also were soon discovered (Van Paradijs
et al. 1997; Frail et al. 1997). These studies showed that the afterglows associated with
GRBs are rapidly fading sources, with X-ray and optical light curves characterised by a power law
decay $\propto t^{-\alpha}$ with $\alpha \div 1-1.5$. Moreover, while 
most of the GRBs, if not all, had an associated X-ray afterglow only about 60\% of them had
also an optical afterglow, i.e. a good fraction of them were dark--GRBs.
For a general review on these topics see Zhang \& Meszaros (2004)
and Piran (2005). 
Therefore, it was clear that to properly study the GRBs,
and in particular the associated afterglows, we needed a fast-reaction
satellite capable of detecting GRBs and of performing immediate multiwavelength
follow-up observations, in particular in the X-ray and optical bands.
{\it Swift} (Gehrels et al. 2004) is designed specifically to study GRBs
and their afterglow in multiple wavebands. It was successfully launched on 2004 November 20, opening
a new era in the study of GRBs (see also N.Gehrels this conference). {\it Swift} has on board three instruments:
a Burst Alert Telescope (BAT) that detects GRBs and determines their positions in the sky
with an accuracy better than 4 arcmin in the band 15-350 keV (Barthelmy et al. 2005a);
an UV-Optical Telescope (UVOT) capable of multifilter photometry with a sensitivity
down to 24$^{th}$ magnitude in white light and a 0.3 arcsec positional accuracy
(Roming et al. 2005); an X-Ray Telescope (XRT) that provides fast X-ray photometry and CCD
spectroscopy in the 0.2-10 keV band with a positional accuracy better than 5 arcsec.
Here we will briefly describe the XRT overall properties, provide
its in-flight performance and outline its main scientific results.
For a more detailed description of the XRT characteristics see Burrows et al. (2004, 2005a).

\section{XRT Description}

A grazing incidence Wolter I telescope provides the XRT imaging capabilities, 
focusing the X-rays onto a CCD at a focal length of 3.5 meters. 
The mirror module is made of 12 nested gold-coated electroformed Ni mirrors.
To prevent temperature gradients that would degrade the image quality,
two thermal baffles in front of the mirror maintain the mirror temperature at 
a constant value of about 20 C. 
The focal plane camera houses an XMM/EPIC MOS CCD, a $600 \times 600$ array 
of $40 \mu$m$\times 40 \mu$m pixels, that corresponds 
to 2.36 arcseconds in the sky. Four calibration sources illuminate the CCD corners 
continuously, allowing us to monitor any CCD response degradation during the
mission life time. A thin Luxel filter mounted in front of the CCD blocks
the optical light. 
A thermo-electric cooling (TEC) system is capable of maintaining
the CCD temperature at -100 C, with the heat dumped to a radiator sitting at a
temperature between -85 C and -45 C, depending on orbital parameters and spacecraft
orientation. The XRT structure is provided by an optical bench interface flange (OBIF)
and a telescope tube, composed of two sections mounted on the OBIF; the forward
telescope tube that supports the star trackers and the telescope door, and the rear tube
that supports the focal plane camera. The mirror module, which is inside the front tube, is mounted
directly on the OBIF through a mirror collar.  Electron deflection magnets are placed behind
the rear face of the OBIF to prevent electrons that pass through the mirror from reaching the detector.
A telescope alignment monitor
provides an accurate measurement of the alignment between the XRT boresight and
the star trackers that are directly mounted on the XRT forward tube
(see Burrows et al. 2004, 2005a).

The main science requirements that drove the design of XRT are: rapid, accurate positions
(better than 5 arcsec in less than 100 seconds after the burst), moderate resolution 
spectroscopy (better than 400 eV at 6 keV after three years of operation) and accurate
photometry for sources spanning up to seven order of magnitude in flux with high time
resolution. A final requirement is that XRT must be able to operate autonomously in order to
provide the afterglow X-ray position very rapidly and efficiently follow the afterglow decay.
To reach these goals, four observing modes have been implemented: an Image Mode (IM),
Photo Diode mode (PD), Window Timing mode (WT) and Photon Counting mode (PC).
In IM the CCD operates like an optical CCD, providing only imaging information
without event recognition (i.e. no spectral capability). This mode is used as soon as
{\it Swift} slews to the position of a new GRB just detected by the BAT. The XRT starts to accumulate
images in IM, looks for the new source and determines its coordinates. These are sent down
via TDRSS and distributed immediately to the community through the GCN network.
Meanwhile, the XRT switches between the other operating modes depending on the
source count rates. The PD mode provides fast timing
resolution of about 0.14 milliseconds, but no imaging information. The entire field of view,
including the corner calibration sources, will end up as a single pixel image. This mode
is used for very bright sources, brighter than 2-3 Crabs.
When the source flux goes below few Crabs, the XRT switches to WT mode, that provides 1.8 ms
time resolution with spatial resolution along one direction (1-D image).
Finally, for sources with fluxes below 1 mCrab, the XRT will operate in PC mode, that 
provides the full imaging and spectroscopic information with a time resolution of 2.5 seconds. 
This is the standard mode in which XRT operates most of the time.
 
\begin{figure}
\centering
\epsfig{file=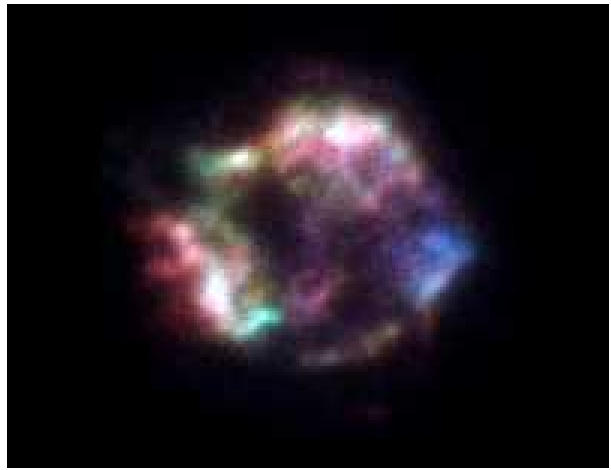,width=\linewidth}
\caption{This colour coded XRT image of the supernova remnant Cas\,A immediately shows
the good image quality that XRT provides.}
\label{fig:casa}
\end{figure}

To establish the XRT performance an end-to-end calibration campaign was performed
in September 2002 at the Panter X-ray calibration facility in Munich. This allowed us
to verify that the XRT Point Spread Function has a HEW of 18 arcsec at
1.5 keV (22 arcsec at 8.1 keV) and it is quite uniform over the entire field of view.
The total effective area of XRT is of $\sim 135 \ {\rm cm^{-2}}$ at 1.5 keV.
The Panter tests also demonstrated that XRT can localise a source with the required 
accuracy and can autonomously and correctly change its readout mode accordingly to
the varying source flux.

\begin{figure}
\centering
\epsfig{file=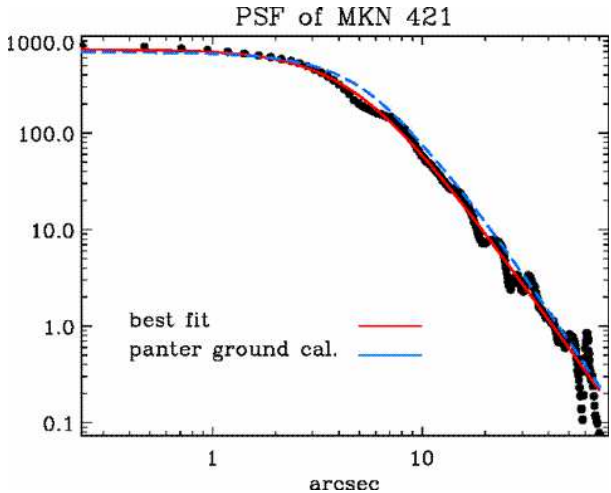,width=\linewidth}
\caption{This figure shows the XRT PSF as measured in-flight on the bright
source MKN\,421. Note how the in flight PSF (continuous line) agrees
perfectly with the one measured on ground (dashed line).}
\label{fig:psf}
\end{figure}

After the {\it Swift} launch, the XRT was turned on 2004 November 23.
However, during the complete activation phase of XRT the TEC system stopped functioning,
leaving the temperature control of the CCD detector only under the passive radiator system. 
Thus, in order to prevent the CCD from getting too hot and loosing sensitivity due to 
increased thermal noise, more stringent pointing
constraints have been implemented to keep the CCD temperature below -50 C (this guarantees
that XRT performs as expected). Since then {\it Swift} has been successfully operated maintaining 
the CCD temperature below this value for most of the time and XRT is delivering very good
quality data with a high degree of observational efficiency (Kennea et al. 2005).
The first light occurred on December 11, when {\it Swift} was pointed to the
bright supernova remnant Cas\,A. XRT provided a superb image that shows structures and filaments
at different temperatures (see Fig. \ref{fig:casa}). After the first
light observation the calibration campaign started and various known X-ray targets were
observed to verify: the PSF as a function of the off-axis angle; effective area;
timing capability; spectral energy resolution and source location capability
(e.g. Hill et al. 2005a; Moretti et al. 2005a, Romano et al. 2005a, Osborne et al. 2005). 
These measurements confirm the perfect functioning of XRT (see e.g. Fig. \ref{fig:psf}).

On May 28, 2005 a micrometeorite shower hit the CCD damaging various pixels. As a result a few
hot columns had to be vetoed otherwise they would saturate the telemetry. This can be
done in WT and PC modes, but not in PD mode. Therefore, the latter mode has not been used 
since then and the XRT, after the IM mode exposures switches to WT mode. 
It is still possibile to recover the PD, by changing the electronic set-up of the focal plane camera.
The balance between the impact on the {\it Swift} operations implied by these changes and
the scientific loss due to the missing PD mode are currently under evaluation
(for a more complete discussion on this topic see T.Abbey this conference).
So far the scientific loss due to the lack of the PD mode seems to be negligible. In any case
we have a few CCD columns that are off and 3 of them are quite in the center of the
field of view, therefore when the source lies on top or near these columns one has
to correct for the sensitivity loss in order to derive the correct source count rate. 
This is now automatically done by the XRT pipeline software distributed by the {\it Swift}
consortium.

\section {XRT observations}

\begin{figure}
\centering
\epsfig{file=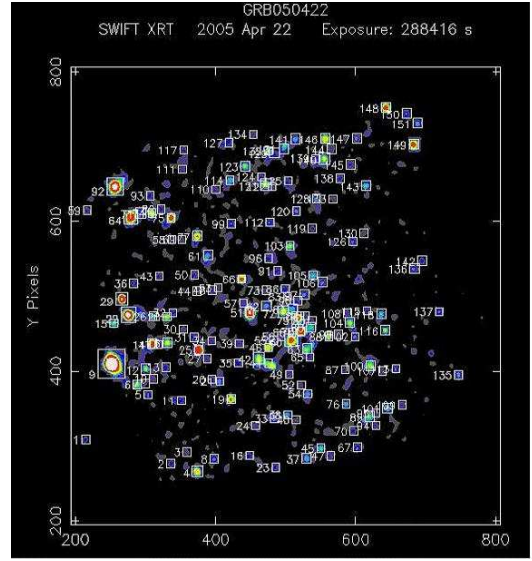,width=1.2\linewidth}
\caption{A 258\,ks XRT exposure on the field of GRB\,050422. 
At the sensitivity limit reached in this exposure
we expect to have about $\sim 1000$ sources/sqdeg. Therefore, hundreds of sources
will be detected in the XRT deep exposures, allowing us to investigate the LogN-LogS of
the X-ray sources making up the cosmic X-ray background.}
\label{fig:deep}
\end{figure}

\begin{figure}
\centering
\epsfig{file=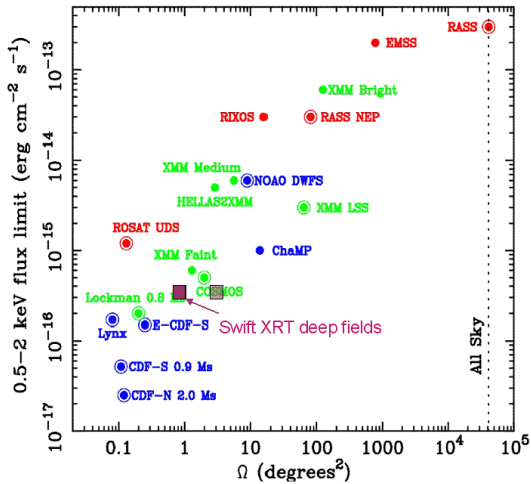,width=1.\linewidth}
\caption{Comparison of the XRT sky coverage at the
sensitivity limit of $\sim 3 \times 10^{-16}$\,erg\,s$^{-1}$\,cm$^{-2}$ in the band 0.5-2.0 keV
with those of other X-ray surveys (adapted from Brandt \& Hasinger 2005). The two XRT points
(filled squares) correspond to the XRT sky coverage calculated at the beginning of September 
2005 and that one estimated after two years of operations.}
\label{fig:survey}
\end{figure}

Due to its low orbit and pointing constraints, a source can not be observed
continuously  by {\it Swift}. Typically three to four targets are observed during each 96
minutes orbit. Depending on its intensity and fading behaviour the X-ray afterglow
of a GRB is monitored with XRT up to a few days--weeks.
Therefore, on some GRBs the XRT total exposure can be as long as few hundred kiloseconds.
This, together with the good PSF of XRT, that is almost flat on the central 8 arcminutes
radius of the field of view, and its low background, due to the low orbit, allow us to obtain
very deep X-ray images (e.g. Fig. \ref{fig:deep}) that can be use to study
the cosmic X-ray background and derive its LogN-LogS. Clearly, we can not go as deep
as the Chandra deep fields, but still can reach a sensitivity limit of
$\sim 3 \times 10^{-16}$\,erg\,cm$^{-2}$\,s$^{-1}$ in the 0.5-2.0 keV band. We estimate that in a two years life time
of the {\it Swift} mission we will cover 2 to 3 square degree of the sky at this limit,
nicely complementing the Chandra and XMM-Newton surveys (Giommi et al. in preparation,
see also Fig. \ref{fig:survey}). Although the goal of XRT is to study the X-ray sources associated 
to GRBs, still this is a nice serendipitous result that XRT will provide.

Up to the end of September 2005, the XRT observed the field of 70 GRBs, always detecting
a X-ray source associated with the GRB, except in 5 cases. Of the 65 X-ray afterglows detected,
41 were detected by XRT within 200\,s of the burst trigger, and 20 in less than 100\,s.
This immediately shows how well {\it Swift} is operating and how efficiently XRT is working.
For the five XRT non-detections, in three cases the XRT observation started 86,  14 and 9 hours
after the burst, respectively. Therefore it is not surprising that we did not found an X-ray afterglow.
In another case the XRT observation started 1.6 hours after the burst. Again this non-detection
is compatible with a weak X-ray afterglow that was not any more detectable.
The last non-detection case corresponds to a short GRB and now we know that the X-ray (and optical)
afterglow associated to the short GRBs are usually weaker than those og long GRBs (see below). Therefore,
we can conclude that so far we always found a X-ray source associated to long GRBs, provided that we 
observe the GRB field on a timescales of a few minutes. This is a further confirmation that the X-ray
observations are the most efficient way to study the afterglows associated to GRBs. The accuracy with
which XRT is localising the X-ray afterglow is of $\sim 4$ arcsec (90\% confidence,
Moretti et al. 2005b). In spite of this good positional accuracy that allow deep follow-up
searches for the optical counterpart, about 50\% of the XRT afterglows are without an optical counterpart. 
This lack of an optical counterpart can either be due to intrinsic absorption in the GRB environment
or to a very high redshift GRBs. We will now outline some of the most exciting results so far achieved
by XRT.

\begin{figure}
\centering
\epsfig{file=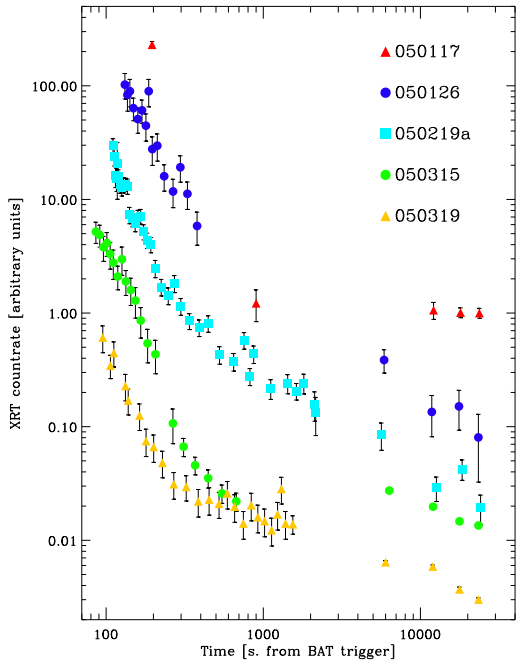,width=0.9\linewidth}
\caption{The steep early X-ray light curves of five GRBs observed by XRT up to March, 2005.
For each GRB the XRT count rates are rescaled by an arbitrary constant factor for clarity.}
\label{fig:steep}
\end{figure}

{\bf The early X-ray light curves of GRBs and afterglows: steep-flat-steep or flat-steep shape.}
The first GRBs observed with XRT is GRB\,041223. In this case {\it Swift} was on target after 4.6 hours,
because the automatic re-pointing of {\it Swift} was still not allowed during this checking phase.
Therefore, a command from ground was issued and the satellite repointed. A new bright and rapidly
fading X-ray source was immediately discovered and soon after also the optical counterpart was
identified (Burrows et al. 2005b).
The automatic slewing of the satellite was enabled on January 2005 with the XRT observations starting
from a few tens to a few hundred seconds after the burst (e.g. Campana et al. 2005, Hill et al. 2005b).
The first unexpected results was the detection of a very steep decay of the X-ray flux 
(${\rm F(t)} \propto {\rm t}^{-\alpha}$ with $\alpha \ga 3$), that breaks to a flatter slope in
the first few hundred seconds 
(Tagliaferri et al. 2005a, see Fig. \ref{fig:steep}). From the lack of a spectral evolution of the X-ray
spectrum of the afterglow across this break, the clear difference of the prompt spectrum measured
by BAT from that one of the afterglow measured by XRT and the discontinuity of the BAT and XRT
light curves, Tagliaferri et al. (2005a) concluded that, at least for the well studied case
of GRB\,050219A, the early steep decay detected by XRT was not due to the prompt emission.
However, a different situation was found for subsequent GRBs. In particular for GRB\,050315
and GRB\,050319, a spectral change across the break, with the spectrum evolving from a softer 
to harder shape, was detected with XRT. In these cases the XRT and BAT
light curves seem to joint smoothly (Cusumano et al. 2005a; Vaughan et al. 2005).
It is now believed that the early steep decay, seen by XRT in most cases, is the tail of the prompt
emission, due to photons emitted at large angle with respect to the observer line of sight
(see also P.O'Brien this conference).
The different behaviour of GRB\,050219A could be explained by the presence of a strong flare 
that it is only partially detected by XRT, that sees only the decaying part. In fact, strong
flares have now been seen by XRT in the early phases of various GRBs (see below).

The study of the X-ray light curves of various GRBs on time scales from a few seconds to
hours-days shows that two common behaviours are emerging, where the light curve consists
of either three or two power law segments (Chincarini et al. 2005; Nousek et al. 2005).
In the first case, which seems to be the most common, there is an initial very steep
decay ($\propto {\rm t}^{-\alpha}$ with $\alpha \ga 3$),
followed by a flattening (with $\alpha \div 0.2-1.0$) and then by a further steepening
($\alpha > 1.0$). The first break occurs in the first one thousand seconds and the second one,
usually before the first 10\,ks. In the second case, the initial very steep part is not seen,
although the observations start few tens of seconds after the burst explosion
(e.g. GRB\,050128, GRB\,050401, GRB\,050525A). In general, during these transitions
the X-ray spectrum remains constant within the observational errors (with just a couple
of exceptions during the early steep decay, see above).
Finally by studying seven bursts for which the redshift was known, Chincarini et al. (2005)
showed that the energy emitted during the afterglow phase correlates with the one emitted
by the prompt and that the afterglow flux emitted in the 0.2-10 keV band goes from few $\%$ up
to $\sim 40\%$ of the flux emitted during the prompt phase in the 15-350 keV band.

\begin{figure}
\centering
\epsfig{file=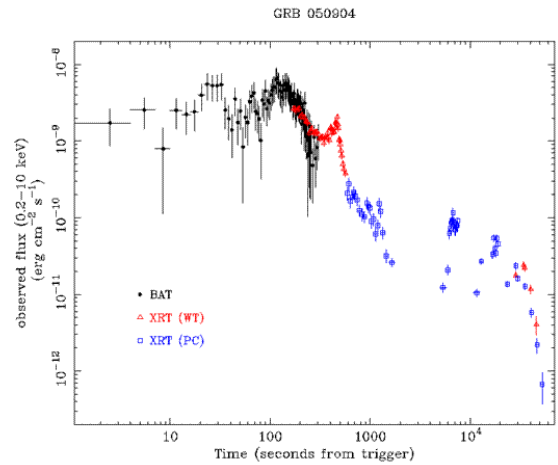,width=\linewidth}
\caption{BAT and XRT light curve of the high redshift GRB\,050904.
Note that the 0.2-10 keV observed flux corresponds to the flux
emitted in the 1.4-73 keV energy band and that the time along the X axis
is also stretched by an amount of (1+z) in the source rest frame,
allowing us to better follow the flares evolution.}
\label{fig:highz}
\end{figure}

{\bf The X-ray flares seen in the XRT light curves.}
On April 6, 2005, a new GRB was detected by BAT and thanks to the {\it Swift} prompt 
slew the XRT started imaging the field around the BAT position 84\,s after the trigger.
The XRT detected a weak decaying X-ray source, that however a few tens of seconds later
started to brighten. Its flux increased by a factor of 6 peaking at 213\,s and 
then start decaying again, the flare ended at $\sim 300$\,s after the burst, then
the X-ray light curve followed again. The power law decay showed before the flare
(Burrows et al. 2005c; Romano et al. 2005b). After this first detection, a very bright
flare was detected in the XRT light curve of GRB\,050502B, with a total fluence that 
exceeded that one of the prompt burst seen by BAT (Burrows et al. 2005c, Falcone et 
al. 2005). The spectra during these and other flares are significantly harder that those
measured before and after the flare, in particular they are harder at the
flare onset and then get softer while the flare evolves. After these first detections,
many other flares have been detected (see also D.Burrows and G.Chincarini this conference).
Here we will only mention two other notable cases, those of GRB\,050724
(Barthelmy et al. 2005b), a short GRB with a very bright X-ray afterglow (see below),
and GRB\,050904, a very high redshift GRB (z=6.29, Kawai et al. 2005; Tagliaferri et al. 2005b)
whose X-ray light curve (see Fig. \ref{fig:highz}) shows strong and complex flare activity (Cusumano et al. 2005b;
Watson et al. 2005). All of the evidence we have so far suggests that these flares,
at least the ones occurring in the first few hundred seconds, are related
to continued internal engine activity.

\begin{figure*}
\centering
\epsfig{file=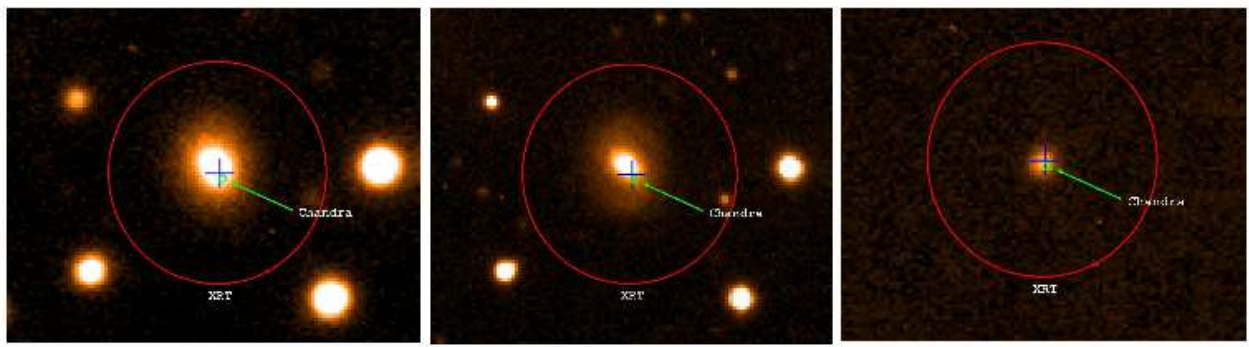,width=1\linewidth}
\caption{VLT optical images showing the position of GRB\,050724, a short burst detected by BAT.
The red circle shows the XRT error box of 6 arcsec radius, the green circle provides the Chandra
position. while the blue cross gives the position of the optical transient (from Barthelmy et al. 2005b).
The left panel shows the VLT image taken on the first night and the panel in the middle shows the VLT
image taken on the second night. The panel on the right shows the result of the image subtraction,
clearly showing the optical transient coincident with the X-ray afterglow.
}
\label{fig:short}
\end{figure*}

{\bf The detection of the first counterpart to the short GRBs.}
The discoveries that have been made in the recent years have established that long 
GRBs probably originate from core-collapse explosion of massive stars. On the other hand
no clues were found on the origin of short GRBs. In fact, while since 1997 astronomers
have been able to study the afterglows associated with long GRBs, for the short GRBs 
this has not been possible up to May 9, 2005, when the BAT detected a short burst, 
GRB\,050509B. This burst was promptly pointed at by {\it Swift}, the XRT started imaging the
field 62\,s after the burst and detected an uncataloged weak X-ray source (11 counts in total)
inside the BAT error circle, providing the first accurate position of a short GRB (Gehrels et 
al. 2005). The X-ray afterglow quickly faded below the detection limit and no optical afterglow 
was detected, in line with the past failure in localising short GRBs. The X-ray afterglow lies
near a luminous non-star-forming elliptical galaxy with z=0.225. 
After the localisation of the short burst GRB\,050709 by HETE-2 (Villasenor et al. 2005), that
lead to the identification of an optical counterpart in a nearby galaxy (z=0.1),
BAT localised another short GRB on July 24. Again {\it Swift} reacted promptly and the XRT started 
to observe the GRB field 74\,s after the burst, this time detecting a very bright and flaring
X-ray afterglow (Barthelmy et al. 2005b). The accurate XRT and subsequent Chandra X-ray position
allowed also to identify the optical (see Fig. \ref{fig:short}) and radio counterpart
(Barthelmy et al. 2005b; Berger et al. 2005). Also in this case the burst is localised off-center
in an elliptical galaxy at z=0.258. These results are consistent with the hypothesis 
that short GRBs originate from the merger of neutron star or black hole binaries. 
Also the isotropic energy emitted in the prompt phase of these short GRBs are 2-3 orders of
magnitude lower than that emitted by the long bursts, again supporting the idea that short
and long GRB have a different origin.

\section{Conclusions}

In conclusion the first ten months of operations have shown that the XRT provides excellent 
quality X-ray images with low background and that the overall calibration is in good shape.
Thanks to this, the XRT is delivering spectacular results, maybe somewhat different from 
those expected. There seem to be two types of X-ray afterglow light curves: steep-flat-steep
(more common) and flat-steep. 
On top of these light curves, various bright flares episodes are detected. 
There is strong evidence that, at least the flares detected in the first few hundred
seconds, are related to continued internal engine activity.
The XRT is providing arcsec localisation for the X-ray counterpart of short burst,
allowing us to investigate the origin and properties of these, so far, elusive sources.

\section*{Acknowledgments}

These activities are supported in Italy by ASI grant I/R/039/04, at Penn State by NASA
contract NAS5-00136 and at the University of Leicester by PPARC grants PPA/G/S/00524 and
PPA/Z/S2003/00507.

\end{document}